\documentclass[onecolumn,showpacs,amsmath,amssymb]{revtex4}
\usepackage{enumerate}
\usepackage{dcolumn}
\usepackage{graphicx}
\usepackage{bm}
\begin{document}
\title{One method to solve Wheeler-DeWitt equation including black hole universe}

\author{Shintaro Sawayama}
 \email{sawayama0410@gmail.com}
\affiliation{Department of Physics, Tokyo Institute of Technology, Oh-Okayama 1-12-1, Meguro-ku, Tokyo 152-8550, Japan \\
and Sawayama Cram School of Physics, Atsuhara 328, Fuji-shi, Shizuoka-ken, Japan}
\begin{abstract}
One of the unsolved issues in the quantum gravity comes from the 
Wheeler-DeWitt equation, which is the second order functional derivative equation with non-linear term.
In this paper, we introduce a method to solve the Wheeler-DeWitt equation with the static restriction introduced in this paper.
Usually one treats the state functional of the spacetimes by the 3-dimensional metrics, which do not contain the timelike metrics.
However we can expand this state to the state which has support on the space of the spacetime metrics with using additional constraint 
which requires the recovery of the 4-dimensional quantum gravity.
Enlarging the support of the state functional of the spacetime metrics, 
we can simply solve the usual Wheeler-DeWitt equation with the additional constraint.
Using this method we can solve some unsolved problems, 
such as the quantization of the black hole.
\end{abstract}

\pacs{04.60.-m, 04.60.Ds}
\maketitle

\section{Introduction}\label{sec1}
The theory of the canonical quantum gravity is based on the Wheeler-DeWitt equation and traces back to 1967 \cite{De}.
There are many methods treating the canonical quantum gravity.
For example Hartle and Hawking \cite{Hart} considered a mini-superspace model.
A loop approach to the canonical quantization was presented by Ashtekar and collaborators \cite{Rov}\cite{As}\cite{Thi}. 
Or there are string quantum cosmology \cite{AL}.
In both schemes the difficulties with the Wheeler-DeWitt equation seams to remain.
And only the homogeneous spacetimes can be quantized,
while inhomogeneous spacetimes can not yet quantized completely.
So we should try to remove or bypass the difficulties of the Wheeler-DeWitt equation. 

In this paper, we show a technical method to solve the Wheeler-DeWitt equation, which we call up-to-down method.
The up-to-down method consists of the following steps. 
First we add another dimension as an external time to the usual 4-dimensional metrics 
and create an enlarged Hilbert space which has support of the spacetime metrics, 
and then we reduce this quantum state to the physical 4-dimensional state, 
we can obtain the additional constraint to simplify the usual Wheeler-DeWitt equation i.e. we use static restriction in a quantum gravity. 
We show how this technique useful to the theories of canonical quantum gravity. 
The same method, however does not work for Klein-Gordon systems.
The strength of our method in the quantum gravity
is shown by considering of the Schwarzschild black hole.
There is no idea of the statics in the quantum gravity.
So our method is strong tool to solve the static Wheeler-DeWitt equation or usual Wheeler-DeWitt equation. 
The consistency of the additional constraint is checked by analyzing solution in the enlarged Hilbert space. 
In this paper, we consider the only vacuum spacetimes. 

In section \ref{sec2}, we introduce the theory of the enlarged Hilbert space with one additional dimension. 
And show how we can solve the Hamiltonian constraint without fixing spacetime.
This analysis is forward to the general discussion of the Wheeler-DeWitt equation.
In subsection \ref{sec21} we introduce what we call the up-to-down method.
In subsection \ref{sec22} we calculate general static solutions of the Wheeler-DeWitt equation.

In section \ref{sec3}, we analyze three mini-superspace models 
as an explicit test of the practicality of the up-to-down method introduced in this paper.
In particular in \ref{sec31}, we consider a Friedmann universe model, and in subsection \ref{sec32} we study an off-orthogonal metric space model, and 
in subsection \ref{sec33}, we quantize a Schwarzschild black hole.
The black hole quantization is the main result achieved using the method introduced in this paper.
Section \ref{sec4} is devoted to a summary and discussion of the main results of our paper.

\section{Up-to-down method and general solutions of the static Wheeler-DeWitt equation}\label{sec2}
\subsection{Up-to-down method}\label{sec21}
In this section we explain in details what we call the up-to-down method. 
We analyze the auxiliary Hilbert space and the original Hilbert space and the projection.

We justify the definition of the projection given in this paper. 
Usually the Hilbert space has the support of the 3-dimensional spacelike metrics. 
However, we expand it here to have the support of the 4-dimensional spacetime metrics.  
 We start by introducing the additional dimension which is an external euclidean time with positive signature, 
and thus create an artificial enlarged Hilbert space corresponding to this external time.
We write such external dimension as $s$.
Although there are many study of the higher dimensional spacetime,
there is no physical meaning of $s$.
However, there is mathematical meaning of the another time.
This idea based on the Isham's quantum category.
So, we imagine many branes or many world interpretation and we quantize many world at the same time. 
The action may be written as
\begin{eqnarray}
S=\int _{M\times s}{}^{(5)}RdMds. \label{f1}
\end{eqnarray}
Where ${}^{(5)}R$ is the 5-dimensional Ricci scalar, 
built from the usual 4-dimensional metric and external time components. 
Rewriting the action by a 4+1 slicing of the 5-dimensional spacetime with lapse functionals given by the $s$ direction, 
we obtain the 4+1 Hamiltonian constraint and the diffeomorphism constraints as,
\begin{eqnarray}
\hat{H}_S\equiv \hat{R}-\hat{K}^2+\hat{K}^{ab}\hat{K}_{ab} \label{f2}\\
\hat{H}_V^a\equiv \hat{\nabla} _b(\hat{K}^{ab}-\hat{K}\hat{g}^{ab}),\label{f3}
\end{eqnarray}
where a hat means 4-dimensional and $a,b$ runs for $0,\cdots , 3$, 
e.g. the $\hat{K}_{ab}$ is extrinsic curvature defined by $\hat{\nabla}_a s_b$ and $\hat{K}$ is its trace, 
while $\hat{R}$ is the 4-dimensional Ricci scalar, 
and $\hat{\nabla} _a$ is the 4-dimensional covariant derivative. \\
 \\
{\it Definition 1. } The artificial enlarged functional space is defined by 
$\hat{H}_S|\Psi^{5} (g)\rangle =\hat{H}_V^a|\Psi^{5} (g)\rangle =0$, 
where $g$ is the 4-dimensional spacetime metrics 
$g_{\mu\nu}$ with ($\mu =0,\cdots ,3$).
We insert the inner product usual sense i.e. $\langle \Psi^{5}(g)^{\dagger}|\Psi^{5}(g)\rangle$.
We write this functional space as ${\cal H}_5$.
 \\ 
 
Here, the definition of the canonical momentum $P$ is different from the usual one. 
Note in fact that the above state in ${\cal H}_5$ is not the usual 5-dimensional quantum gravity state, because the 4+1 slicing is along the $s$ direction.
This is why we call this Hilbert space as artificial functional space. 
It is not defined by $\partial {\cal L}/(\partial dg/dt)$ but by $\partial {\cal L}/(\partial dg/ds)$, where 
${\cal L}$ is the 5-dimensional Lagrangian. 

In addition, we impose that 4-dimensional quantum gravity must be recovered from the above 5-dimensional action.
The 3+1 Hamiltonian constraint and diffeomorphism constraint are,  
\begin{eqnarray}
H_S\equiv {\cal R}+K^2-K^{ij}K_{ij} \label{f4}\\
H_V^a\equiv D_j(K^{ij}-Kq^{ij}).\label{f5}
\end{eqnarray}
where $i,j$ runs for $1,\cdots ,3$.
Here $K_{ab}$ is the usual extrinsic curvature defined by $D_at_b$ 
and $K$ is its trace, while ${\cal R}$ is the 3-dimensional Ricci scalar, 
and $D_a$ is the 3-dimensional covariant derivative.
Then we can define a subset of the auxiliary Hilbert space on which the wave functional satisfies the usual 4-dimensional constraints. 
In order to relate the 4 and 5 dimensional spaces we should define projections.
\\ \\
{\it Definition 2.} The subset of ${\cal H}_5$ in which the five dimensional quantum state satisfies the
extra constraints $H_S\Pi^1|\Psi ^5(g)\rangle=H_V^a\Pi^1|\Psi ^5(g)\rangle =0$
is called ${\cal H}_{5lim}$, where $P$ is the projection
defined by
\begin{eqnarray}
\Pi^1 :{\cal H}_5 \to {\cal F}_4 \ \ \ 
\{ \Pi^1 |\Psi^5(g)\rangle=|\Psi^5(g_{0\mu}={\rm const})\rangle \} ,\label{f6}
\end{eqnarray}
where ${\cal F}_4$ is a functional space.
And ${\cal H}_4$ is the usual four dimensional quantum gravity state with the restriction that 
$H_S|\Psi^4(q)\rangle=H_V^a|\Psi ^4(q)\rangle=0$.
Here $q$ stands for the 3-dimensional metric $q_{ij}(i=1,\cdots ,3)$, and
$\Pi^{2}$ is defined by
\begin{eqnarray}
\Pi^{2}:{\cal H}_{5lim} \to {\cal H}_4 .\label{f7}
\end{eqnarray}
\\

Although the measure of the projection may be zero, we can justify this projection in the next subsection.

From now we consider the recovery of the 4-dimensional vacuum Einstein gravity from the 5-dimensional wave functional.
We consider 4-dimensional Ricci scalar in the 4+1 Hamiltonian constraint.
The 4-dimensional Ricci scalar is decomposed as
\begin{eqnarray}
\hat{R}=H_S+n_aH_V^a-\frac{1}{2}\dot{P}.\label{f8}
\end{eqnarray}
Here $n_a$ is the orthonormal vector of the time and $P$ is the contraction of momentum. 
Then the modified Hamiltonian constraint for the 5-dimensional 
quantum state which contains the 4-dimensional Einstein gravity becomes,
\begin{eqnarray}
\hat{H}_S\to -m\hat{H}_S:= -\hat{K}^2+\hat{K}^{ab}\hat{K}_{ab}-\frac{1}{2}\dot{P} \label{f9}
\end{eqnarray}
There is the theoretical branch in using the Dirac constraint or Hamiltonian and diffeomorphism constraint.
To use the Dirac constraint at this point create the additional constraint which restrict the state to the static.

Finally, the simplified Hamiltonian constraint in terms of the canonical representation becomes
\begin{eqnarray}
m\hat{H}_S =(-g_{ab}g_{cd}+g_{ac}g_{bd})\hat{P}^{ab}\hat{P}^{cd}-\frac{1}{2}\dot{P}. \label{f10}
\end{eqnarray}
The magic constant factor $-1$ for the term $g_{ab}g_{cd}$ is a consequence of the choice of dimensions for ${\cal H}_5,{\cal H}_4$.
Here $\hat{P}^{ab}$ is the canonical momentum of the 4-dimensional metric 
$g_{ab}$, that is $\hat{P}^{ab}=\hat{K}^{ab}-g^{ab}\hat{K}$. 
And as we mentioned above, this canonical momentum is defined by the external time and not by the usual time.

We now give a more detailed definition of the artificial functional space as follows: \\ \\ 
{\it Definition 3.} The subset ${\cal H}_{5(4)}\subset {\cal H}_5$ is defined by the constraints,  
$ \hat{R}|\Psi ^5(g)\rangle =0$, 
and we write its elements as $|\Psi ^{5(4)}(g)\rangle$.
We also define a projection $\Pi^3$ as
\begin{eqnarray}
\Pi^3 : {\cal H}_{5(4)} \to {\cal H}_{4(5)} \ \ \  \{ P^*|\Psi ^{5(4)}(g)\rangle =|\Psi ^{5(4)}(g_{0\mu}={\rm const})\rangle
=: |\Psi ^{4(5)}(q)\rangle \} , \label{f11}
\end{eqnarray} 
where ${\cal H}_4$ is a subset of ${\cal H}_{4(5)}$.
We can defien the inner product in the ${\cal H}_{4(5)}$ space like, $\langle \Psi^{4(5)}(q)^{\dagger}|\Psi^{4(5)}(q)\rangle$
\\ 

We notice the projection $\Pi^1$ and $P^2$ and $\Pi^3$ is all most all same. 
However, we use the different symbol because the projected functional space is different.\\ 

{\it Definition 4.} 
In ${\cal H}_{4(5)}$ there is a subset whose state satisfy $H_S|\Psi^{4(5)}(q)\rangle=H_V|\Psi ^{4(5)}(q)\rangle=0$. 
We write such Hilbert 
space as ${\cal H}_{4(5)phys}$ and its elements as $|\Psi_{phys}^{4(5)}(q)\rangle$. 
If there are relations $\Pi^3|\Psi^{5(4)}(g)\rangle=|\Psi_{phys}^{4(5)}(q)\rangle$, we write such 
$|\Psi^{5(4)}(g)\rangle$ as $|\Psi^{5(4)}_{phys}(g)\rangle$ 
and we write such Hilbert space ${\cal H}_{5(4)phys}$.
\\ 

The ${\cal F}_4$ functional space or ${\cal H}_{5(4)}$ space does not need to be the $l^2$ norm spaces.
As a consequence, we can obtain following theorem: \\  \\

We comment on the projections.
Although there are three projection, these operation is same.
However, the projecting space and projected space is different.
So we use the number 1,2,3.
The projection act only on the right hand side, it does not act left hand side.
And usually it does not commute with the Hamiltonian constraint and the diffeomorphism constraint.
Although the inverse projection like $\Pi^{-1}$ may be one to many projection,
we can choose suitable inverse projection so that the measure of the projection does not become zero.
Or we assume such projection exist.
Although the measure of the projection may be zero, we assume there is at least one enlargement whose measure of the projection is not zero.
Or we can only treat such subspace.
\\

{\it Theorem 1.} In this method, in the ${\cal H}_4$ additional constraint $m\hat{H}_S\Pi^3=0$ appears which we call static restriction, if there is no time evolution and the projection is defined by the definition 3.
\\

{\it Sketch of the proof}\\
\begin{eqnarray}
\hat{H}_S\Pi^3=\hat{R}\Pi^3-\hat{K}^2\Pi^3 +\hat{K}^{ab}\hat{K}_{ab}\Pi^3 \nonumber \\
\to H_S\Pi^3 +n_aH_V^a\Pi^3-\frac{1}{2}\dot{P}\Pi^3-\hat{K}^2\Pi^3 +\hat{K}^{ab}\hat{K}_{ab}\Pi^3 \nonumber \\
\to -\frac{1}{2}\dot{P}\Pi^3-\hat{K}^2\Pi^3 +\hat{K}^{ab}\hat{K}_{ab}\Pi^3 \nonumber \\
\to (-q_{ij}g_{kl}+g_{ik}g_{jl})P^{ij}P^{kl}-\frac{1}{2}\dot{P}. \label{f12}
\end{eqnarray}
\\

{\it Lemma 1}.If the term of the $\dot{P}$ became to zero acted on the state,
the additional constraint become static restriction as,
\begin{eqnarray}
{\cal S}=(-q_{ij}g_{kl}+g_{ik}g_{jl})P^{ij}P^{kl} \label{f13}
\end{eqnarray}
\\ 
We explain this lemma.
The static restriction $S$ is the special case of the additional constraint and the additional constraint usually does not commute with the Hamiltonian constraint and the diffeomorophis constraints.
However, in the special mini-superspaces the static constraint and the Hamiltonian constraint does commute.
There is another method of the up-to-down method, which uses $m\hat{H}_S$ in the enlarged functional space.
Then there does not appear the additional constraint in the usual 4-dimensional constraints.
This another method is useful to solve the usual static 4-dimensional Wheeler-DeWitt equation.
We only consider static restriction or enlargement of the static restriction in this paper.

Now we also have the additional constraint equation coming from the assumption that the 5-dimensional 
vacuum Einstein gravity should reproduce 4-dimensional Einstein gravity in the classical limit. 
That is,
\begin{eqnarray}
{}^{(4)}G_{ab}=\hat{K}\hat{K}_{ab}-\hat{K}_{a}^{\ c}\hat{K}_{bc}-2\nabla_a\beta_b=0, \label{f14}
\end{eqnarray}
where,
\begin{eqnarray}
\beta ^a:=s^b\nabla _bs^a-s^a\nabla _bs^b \label{f15}
\end{eqnarray} 
If we assume that l.h.s. of Eq. (\ref{f14}) corresponds to the matter term, we can take its trace.
And this additional constraint reduces to the sum of the  4+1 Hamiltonian constraint and the diffeomorphism constraint, that is,
\begin{eqnarray}
8\pi T^a_a:=\hat{K}^2-\hat{K}_{ab}\hat{K}^{ab}
-2\nabla_a\beta^a = m\hat{H}_S-2s_a\hat{H}_V^a-\frac{1}{2}\dot{P} \approx m\hat{H}_S. \label{f16}
\end{eqnarray}
In other words, the matter term $T_a^a$ has been promoted to the operator, 
it does not produce further constraints other than 5-dimensional modified Hamiltonian constraint. 
We don't assume equation (\ref{f14}), because it is too strong, determine the four 
independent metrics by the other metric components. \\ \\ 
{\it Lemma2.} The requirement to recover four dimensional gravity, 
$8\pi T_a^a |\Psi ^5(g)\rangle = 0$, is the same as 
$\hat{H}_S|\Psi ^5(g)\rangle = 0$.  
So $8\pi T_a^a \approx m\hat{H}_S \approx \hat{H}_S$. 

We can summarize the procedure to solve the usual Wheeler-DeWitt equation. \\ \\
{\it Steps to solve the Wheeler-DeWitt equation.} \\
There are 2 steps to solve the usual 3+1 Hamiltonian constraint.\\
a1) Solve $m\hat{H}_S|\Psi^5(g)\rangle =0$ and obtain $|\Psi ^{5(4)}(g)\rangle$.\\
a2) Project this state by $\Pi^3$.\\
a3) Solve $H_S|\Psi ^{4(5)}(g)\rangle =0$ to obtain $|\Psi ^{4(5)}_{phys}(q)\rangle$.\\ \\
b1) Using the additional constraint $m\hat{H}_S\Pi^3|\Psi ^{4(5)}(q)\rangle =0$ in ${\cal H}_4$, 
solve $H_S|\Psi^{4(5)}(q)\rangle =0$ to obtain $|\Psi^{4(5)}_{phys}(q)\rangle$.\\
b2) Then enlarge $|\Psi^{4(5)}_{phys}(q)\rangle \to |\Psi^{5(4)}_{phys}(g)\rangle$. \\ \\
\begin{figure}
\includegraphics{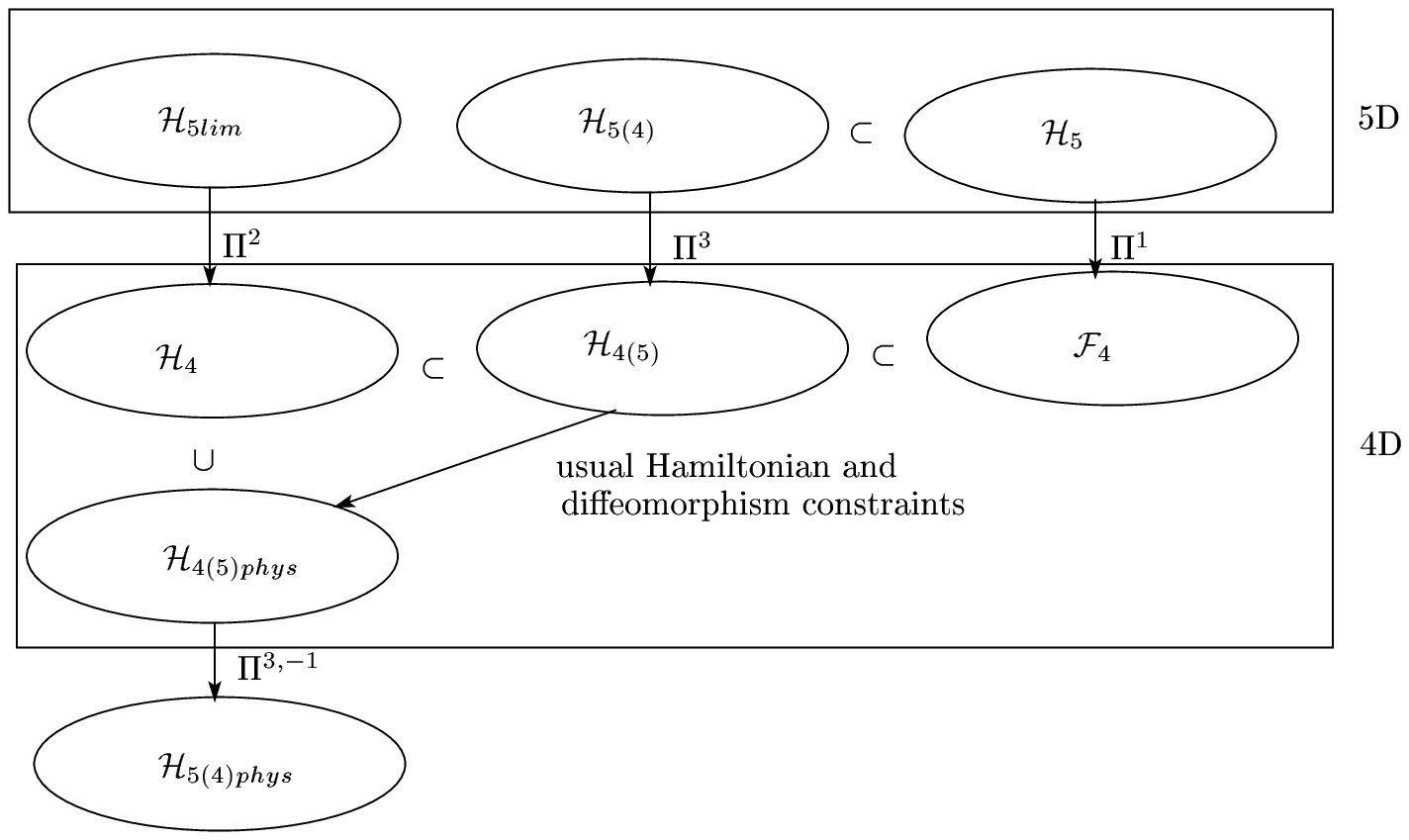}
\caption{The relations between the functional spaces and the physical 4-dimensional Hilbert space.
The Hilbert space ${\cal H}_{4(5)phys}$ is always subset of the physical Hilbert space ${\cal H}_4$.
The $\Pi^1$ and $\Pi^2$ and $\Pi^3$ are same operation. However, projected space is different.
The inverse projection $\Pi^3$ can be defined by one-to-many way.
Although it is mathematically not well-defined, if we can find at least one enlargement, this method works correct way.}
\end{figure}
The step a) is useful when we consider the Hamiltonian constraint in the general sense.
The step b) is useful when we consider mini-superspace models and we use it in section \ref{sec3}

\subsection{General solution for the $|\Psi^{5(4)}(g)\rangle$ state and the projection theorem}\label{sec22}
Let us consider $|\Psi^{5(4)}(g)\rangle$ as the state with support on 
the spacetime metrics which contain 4-dimensional gravity, 
and write the Hamiltonian and diffeomorphism constraints in the operator representation. 
Acting in the auxiliary Hilbert state as,
\begin{eqnarray}
\sum (g_{ab}g_{cd}-g_{ac}g_{bd})
\frac{\delta}{\delta g_{ab}}\frac{\delta}{\delta g_{cd}}|\Psi^{5(4)}(g)\rangle =0 \label{f17}
\end{eqnarray}
\begin{eqnarray}
(\hat{\nabla} _{a}\hat{P}^{ab})|\Psi\rangle 
=\bigg[ \frac{\partial }{\partial x^{a}},\frac{\delta}{\delta g_{ab}}\bigg] |\Psi^{5(4)}(g)\rangle =0. \label{f18}
\end{eqnarray}
We can easily find a solution to Eq. (\ref{f17}), and such solution is the 
superposition of the two following states
\begin{eqnarray}
|\Psi^{5(4)} (g)\rangle _d=\sum_{\mu}f_{(\mu )}[g_{\mu\mu}], \label{f19}
\end{eqnarray}
and
\begin{eqnarray}
|\Psi^{5(4)} (g)\rangle _{od}=\sum_{\mu \geq\nu}f_{(\mu\nu)}^{(1)}[g_{\mu\nu}]. \label{f20}
\end{eqnarray}
Where $d$ stands for diagonal part, and $od$ stands for off-diagonal part. 
Here $f_{(\mu )}$ is an arbitrary functional of only the $g_{\mu\mu}$ and 
$f_{(\mu\nu)}^{(1)}$ is a first order functional of only the $g_{\mu\nu}$.   
This division of the state of the $|\Psi^{5(4)}(g)\rangle$ by $g_{\mu\mu}$ is important. 
The functional $f_{(\mu )}[g_{\mu\mu}]$ can be expressed as a function in the following way,
\begin{eqnarray}
f_{(\mu )}[g_{\mu\mu}]=f_{(\mu )}(g_{\mu\mu},g_{\mu\mu ,\nu},...).\label{f21}
\end{eqnarray}
The diffeomorphism constraint can be simply written as,
\begin{eqnarray}
 \bigg[ \frac{\partial }{\partial x^{\mu}},\frac{\delta}{\delta g_{\mu\mu}}\bigg]f_{(\mu )}[g_{\mu\mu}]
+\sum_{\mu\not=\nu}\bigg[ \frac{\partial }{\partial x^{\nu}},\frac{\delta}{\delta g_{\mu\nu}}\bigg]
f_{\mu\nu}^{(1)}[g_{\mu\nu}]=0
\ \ \ (\mu =0,1,2,3), \label{f22}
\end{eqnarray}
One of the solutions of this equation is,
\begin{eqnarray}
f_{(\mu )}(g_{\mu\mu})\subset f_{(\mu )}[g_{\mu\mu}] ,\label{f23}
\end{eqnarray}
and 
\begin{eqnarray}
0 \subset f_{(\mu\nu )}^{(1)}[g_{\mu\nu}]\ \ \ (\mu\not=\nu ), \label{f24}
\end{eqnarray}
These functionals are one of the solutions for the $|\Psi^{5(4)}(g)\rangle$ state. 
We note that the metric can be locally diagonalized. \\
\quad There are also solutions to the modified Hamiltonian constraint equation, that is the deterministic function as, 
\begin{eqnarray}
|\Psi^{5(4)} (g)\rangle =\sum_{\sigma}\prod _{\mu =0}^3g_{\mu\sigma (\mu )}, \label{f25}
\end{eqnarray}
$\sigma$ stands for permutation.
We note that Eq. (\ref{f25}) does not involve the signature. 
In the derivation of the above formula, we did not assume any symmetry for the metric, in particular $g_{\mu\nu}=g_{\nu\mu}$ was not assumed. 
If we assume such a symmetry, we should add constant factors for the each element in Eq. (\ref{f25}). 
Power series of $g_{\mu\sigma (\mu )} $ are also solutions, we may write
\begin{eqnarray}
|\Psi^{5(4)} (g)\rangle =\sum_{\sigma}\prod _{\mu =0}^3g_{\mu\sigma (\mu )}^n, \label{f26}
\end{eqnarray}
and also functionals of $g_{\mu\sigma (\mu )}$ is also solutions, that is
\begin{eqnarray}
|\Psi^{5(4)} (g)\rangle =\sum_{\sigma}f\bigg[\prod _{\mu =0}^3g_{\mu\sigma (\mu )}\bigg] . \label{26}
\end{eqnarray}
We can make the state (\ref{21}) become a functional of the determinant of $g$ only, 
if we choose the coordinates in such a way that the metric becomes diagonal. \\

{\it Theorem 2.} The projection of the above two states (\ref{19}-\ref{20}), (\ref{26}) $\Pi^3|\Psi^{5(4)}(g)\rangle$ satisfy 
${\cal S}\Pi^3|\Psi^{5(4)}(q)\rangle =0$. 
\\

This fact is one of the motivations for introducing $\Pi^3$. 
Otherwise, this additional constraint does not appear in general solution, if it acts mini-superspaces, additional constraint appears.

\section{Mini-superspaces}\label{sec3}
In this section we consider three mini-superspace models, i.e a Friedmann model, an anti-orthogonal model, and a Schwarzschild model.
In subsection \ref{sec31} we check whether the up-to-down method introduced in section \ref{sec2} is really applicable. 
In subsection \ref{sec32} we study the off-orthogonal metric model.
One main progress is shown in subsection \ref{sec33}, which deals with a Schwarzschild black hole,
and where it is shown that the black hole is quantized by up-to-down method.
In this section we always ignore so called operator ordering.

\subsection{Friedmann universe and the problem of time}\label{sec31}
Because the functional space ${\cal H}_{4(5)phys}$ may be empty, we should check that the up-to-down method 
is really consistent.
In this paper we ignore the operator ordering for simplicity.

We simply assume that the metric corresponding to a Friedmann model with a cosmological constant $\Lambda$, i.e.
\begin{eqnarray}
g_{ab}:= \begin{pmatrix}
b & 0 & 0 & 0 \\
0 & a & 0 & 0 \\
0 & 0 & a & 0 \\
0 & 0 & 0 & a
\end{pmatrix}.\label{f27}
\end{eqnarray}
Here $a,b$ only depend on time $t$, and coordinates are chosen as $t,x,y,z$. 
The cosmological constant is included in the constraint of 4-dimensional gravity. 
The modified Hamiltonian constraint $m\hat{H}_S$ is explicitly written as,
\begin{eqnarray}
m\hat{H}_S=6ab\frac{\partial}{\partial a}\frac{\partial}{\partial b}+a^2\frac{\partial^2}{\partial a^2} \\
=6\frac{\partial}{\partial \eta}\frac{\partial}{\partial \eta_b}+\frac{\partial^2}{\partial \eta^2}=0 ,\label{f28}
\end{eqnarray}
where we have defined $\eta =\ln a$, $\eta _b=\ln b$.
This constraint acts on $|\Psi^5(g)\rangle$, and solution of 
$m{\hat H}_S|\Psi^5(g)\rangle =0$ give $|\Psi^{5(4)}(g)\rangle$. 

The usual 3+1 Hamiltonian constraint is
\begin{eqnarray}
H_S
=\frac{9}{2}a^2\frac{\partial}{\partial a^2}+\Lambda \\
=\frac{9}{2}\frac{\partial}{\partial \eta^2}+\Lambda =0. \label{f29}
\end{eqnarray}
In this Friedmann model we do not need additional constraint $m\hat{H}_SP^* \approx 0$.
If the cosmological constant is zero, ${\cal H}_4$ and ${\cal H}_{4(5)}$ are the same Hilbert space. 
Here we enlarge the ${\cal H}_4$ state $|\Psi^4(g)\rangle $ to the ${\cal H}_{5(4)}$ state.

We can obtain the $|\Psi^4(q)\rangle$ state as,
\begin{eqnarray}
|\Psi^4(\eta )\rangle =\exp (i\frac{\sqrt{2}}{3}\Lambda^{1/2} \eta) . \label{f30}
\end{eqnarray}

The next step is to check the consistency of the up-to-down method.
We can expand the state (21) to the state $|\Psi^{5(4)}(g)\rangle$ by the inverse projection $(\Pi^3)^-1$ as, 
\begin{eqnarray}
|\Psi^{5(4)}(\eta_b,\eta )\rangle =f(\eta_b) \exp (i\frac{\sqrt{2}}{3}\Lambda^{1/2} \eta) .\label{f31}
\end{eqnarray}
Although the inverse projection can be defined another way, if we can find at least one enlargement, this method works well.
Then we can solve the following constraint equation to derive $|\Psi^{5(4)}(g)\rangle$.
\begin{eqnarray}
m\hat{H}_S|\Psi^5 (g)\rangle =\bigg( 6i\frac{\sqrt{2}}{3}\Lambda^{1/2} f'(\eta_b ) -6\frac{2}{9}\Lambda f(\eta_b) \bigg)
\exp (i\frac{\sqrt{2}}{3}\Lambda^{1/2} \eta) =0 \label{f32}
\end{eqnarray}
From this equation we obtain
\begin{eqnarray}
2i\sqrt{2}\Lambda^{1/2} f'(\eta_b ) -\frac{4}{3}\Lambda f(\eta_b) =0, \label{f33}
\end{eqnarray}
or
\begin{eqnarray}
f'=-i\frac{\sqrt{2}}{3}\Lambda^{1/2} f, \label{f34}
\end{eqnarray}
whose solution is
\begin{eqnarray}
f(\eta_b)=\exp \bigg( -i\frac{\sqrt{2}}{3}\Lambda ^{1/2}\eta_b \bigg) . \label{f35}
\end{eqnarray}
From this result and Eq. (\ref{f30}) we can obtain $|\Psi^{5(4)}(g)\rangle$ state as
\begin{eqnarray}
|\Psi^{5(4)} (\eta_b ,\eta)\rangle =\exp \bigg( -i\frac{\sqrt{2}}{3}\Lambda ^{1/2}\eta_b \bigg)\exp (i\frac{\sqrt{2}}{3}\Lambda^{1/2} \eta) .\label{f36}
\end{eqnarray}
This state is a solution to both the $m{\hat H}_S$ constraint and $H_S$ constraint, 
which shows that the up-to-down method is applicable at least to the simple Friedmann model. 
We comment on this enlargement.
The norm of enlarged state is not important.
Although, there are other enlargement, we choose multiplication enlargement.
Because all the projection produces same state without constant factor.
By this enlargement the measure of the projection does not become zero.
\subsection{Quantization of the off-orthogonal metric space}\label{sec32}
As the next example we consider the quantization of a spacetime with anti-orthogonal metric i.e.
\begin{eqnarray}
g_{ab}= \begin{pmatrix}
-c & 0 & 0 & 0 \\
0 & a & b & 0 \\
0 & b & a & 0 \\
0 & 0 & 0 & a
\end{pmatrix}.\label{f37}
\end{eqnarray}
Here $a,b,c$ only depend on time $t$, and the coordinates are chosen as $t,x,y,z$.
The modified 4+1 Hamiltonian constraint is now written as
\begin{eqnarray}
m\hat{H}_S=-6ac\frac{\partial}{\partial a}\frac{\partial}{\partial c}
+(6a^2-2b^2)\frac{\partial^2}{\partial a^2}+2(b^2-a^2)\frac{\partial^2}{\partial b^2}=0, \label{f38}
\end{eqnarray}
while the usual 3+1 Hamiltonian constraint is written as
\begin{eqnarray}
H_S=-(5a^2-b^2)\frac{\partial}{\partial a^2}-(2b^2-a^2)\frac{\partial^2}{\partial b^2}=0.\label{f39}
\end{eqnarray}
Since finding a solution to (\ref{39}) is something difficult, we use additional constraint
\begin{eqnarray}
{\cal S} =(6a^2-2b^2)\frac{\partial^2}{\partial a^2}+2(b^2-a^2)\frac{\partial^2}{\partial b^2}=0, \label{f40}
\end{eqnarray}
to simplify it as
\begin{eqnarray}
H_S=-\frac{(a^2+b^2)(2a^2-b^2)}{(a^2-b^2)}\frac{\partial^2}{\partial a^2}=0.\label{f41}
\end{eqnarray}
Both constraints $H_S$ and $8\pi T_a^aP^*$ act on the state $|\Psi^{4(5)}(q)\rangle$ which belong to a subset of ${\cal H}_4$.
The differential equation $H_S|\Psi^{4(5)}(q)\rangle$ can be solved easily and the result is,
\begin{eqnarray}
|\Psi^{4(5)} (a,b)\rangle =C_1(b)a+C_2(b),\label{f42}
\end{eqnarray}
here $C_1(b),C_2(b)$ are the arbitrary functions of $b$.
The solution which also satisfy $m\hat{H}_SP^*\approx 0$ is now,
\begin{eqnarray}
|\Psi^{4(5)} (a,b)\rangle =E_1ab+E_2a+E_3b+E_4.\label{f43}
\end{eqnarray}
Here $E_1\cdots E_4$ are constants. 
We can finally expand the state (\ref{43}) to the $|\Psi^{5(4)}(g)\rangle$ state, 
using the modified Hamiltonian constraint $m\hat{H}_S$, as
\begin{eqnarray}
|\Psi ^{5(4)} (a,b,c)\rangle =F_1ab+F_2a+F_3bc+F_4c+F_5. \label{f44}
\end{eqnarray}
This enlargement is also consistent with projection measure.
The projected state is $|\Psi^{4(5)}$ state.
In this example the static restriction and the Hamiltonian constraint does not commute.
However, the obtained 4-dimensional state is state of the usual Hamiltonian constraint.

\subsection{Quantization of the Schwarzschild black hole}\label{sec33}
The most interesting result is the applicability of our up-to-down method to the model of a black hole spacetime.
We consider the Schwarzschild black hole metric, i.e.
\begin{eqnarray}
g_{ab}=\begin{pmatrix}
-f & 0 & 0 & 0 \\
0 & f^{-1} & 0 & \\
0 & 0 & A & 0 \\
0 & 0 & 0 & A\sin ^2\theta
\end{pmatrix}.\label{f45}
\end{eqnarray}
Here we choose the usual coordinates $t,r,\theta,\phi$, and $f$ is the 
function that depends only on time $t$, and $A$ stands for the area i.e. $A\doteq r^2$.
The derivative of its components is written as,
\begin{eqnarray}
\frac{\delta}{\delta g_{tt}}=-\frac{\partial}{\partial f} \label{f46} \\
\frac{\delta}{\delta g_{rr}}=-f^2\frac{\partial}{\partial f} \label{f47} \\
\frac{\delta}{\delta g_{\theta\theta}}=\frac{\partial}{\partial A} \label{f48} \\
\frac{\delta}{\delta g_{\phi\phi}}
=\frac{1}{\sin^2\theta}\frac{\partial }{\partial A}+\frac{1}{A\sin 2\theta}\frac{\partial}{\partial \theta}.\label{f49}
\end{eqnarray}
We make the simplifying assumption that the quantum state does not depend on $\theta$, because of spherical symmetry. 
Using the above formulas for the derivatives we can write the modified 4+1 Hamiltonian constraint as
\begin{eqnarray}
m\hat{H}_S=-f^2\frac{\partial^2}{\partial f^2}+A^2\frac{\partial^2 }{\partial A^2}=0, \label{f50}
\end{eqnarray}
while the usual 3+1 Hamiltonian constraint is
\begin{eqnarray}
H_S=\frac{1}{2}\bigg(-f^2\frac{\partial^2 }{\partial f^2}
+Af\frac{\partial }{\partial f}\frac{\partial}{\partial A}
-2A^2\frac{\partial^2 }{\partial A^2}\bigg)+{\cal R}=0, \label{f51}
\end{eqnarray}
where {\cal R} is 3-dimensional Ricci scalar ${\cal R}=M/r^3=(1-f)/A$. 
The equation (\ref{f51}) is a second order partial differential derivative with non-linear terms, 
and it is difficult to find explicit analytical solution.
However, if we use the additional constraint
\begin{eqnarray}
{\cal S}=-2fA\frac{\partial}{\partial f}\frac{\partial}{\partial A}+A^2\frac{\partial^2}{\partial A^2} \\
=A\frac{\partial}{\partial A}\bigg( -2f\frac{\partial}{\partial f}+A\frac{\partial}{\partial A}\bigg) =0, \label{f52}
\end{eqnarray}
Then we can find the following relation between the two parameters $f$ and $A$,
\begin{eqnarray}
f^{1/2}=cA,\label{f53}
\end{eqnarray}
where $c$ is a non-zero complex constant.
At this point we comment on the theoretical branch in Eq.(\ref{f52}).
And only if the above parameter relation holds, we can carry on simultaneous quantization.
There are another way to quantize static black holes, i.e. Eq.(\ref{f51}).
We can find another way which is to use only the differential relation or to determine the form of $f$ in terms of $A$.
Then there appears duality between $A$ and $f$.
The singularity at the $A=0$ and $f=0$ become degenerate. 
And making it parameter relation, we can commute the static restriction and the Hamiltonian constraint. 
From Eq. (\ref{f53}), we can deduce the analytic form of the mass operator, as
\begin{eqnarray}
2\hat{M}=A^{1/2}-c^2A^{5/2}=r-c^2r^5, \label{f54}
\end{eqnarray}
if we assume $f=1-2\hat{M}/r$.
This is what we know from the additional constraint.
Then the mass corresponding to the classical gravity can be defined by,
\begin{eqnarray}
\langle M(c)\rangle :&=&\int_0^{\infty} 2\pi A^{1/2}\langle \Psi |\hat{M}|\Psi\rangle dA \nonumber \\
&=&\langle A\rangle -c^2\langle A^3\rangle .\label{f55}
\end{eqnarray}
For the classical correspondence, we require the averaged mass is real and bounded from above.
From the fact that averaged mass is real, we know $c$ is real or purely imaginary. 
Otherwise the relation of $A$ and $f$ seems to violating asymptotic flatness, if averaged value of 
the mass is constant (does not depend $f$ or $A$) 
and does not diverge, the quantum state is corresponding to the usual classical Schwarzschild black hole. 
The fact that the classical mass (\ref{f55}) is constant is clear from the above formula.

Using the relation of $A$ and $f$ we can rewrite the usual 3+1 Hamiltonian constraint simply
as second order ordinal differential equation, as
\begin{eqnarray}
H_S=-\frac{7}{8}A^2\frac{\partial^2}{\partial A^2}+\frac{1-c^2A^2}{2A}=0. \label{f56}
\end{eqnarray}
Here, we used
\begin{eqnarray}
{\cal R}=\frac{\hat{M}}{r^3}=\frac{1-c^2A^2}{2A} \label{f57}
\end{eqnarray}
This kind of constraint equation (\ref{f56}) can always be solved numerically.
And symbolically, we write its solution as $|\Psi^{4(5)}(c,A)\rangle =C_1F(c,A)$, 
where $C_1$ is constant factor for the later convenience.

If $A\to 0$, the 3+1 Hamiltonian constraint becomes
\begin{eqnarray}
\lim_{A\to 0}H_S&=&-\frac{7}{8}A^2\frac{\partial^2}{\partial A^2}+\frac{1}{2A} \\
&\to &-\frac{7}{16}\frac{\partial^2}{\partial a}+1=0, \label{f58}
\end{eqnarray} 
where, $a=A^{-1/2}$.
Such coordinate transformation can be done because of ignorance of operator ordering.
So, if $A\to 0$, the constant $c$ disappears from the state. 

In this limit the differential equation for $a$ can be solved analytically, and the solution is,
\begin{eqnarray}
|\Psi (c,A\to 0)\rangle =E_1\exp \bigg( \frac{4}{\sqrt{7}}A^{-1/2}\bigg) +E_2\exp \bigg( -\frac{4}{\sqrt{7}}A^{-1/2}\bigg) ,\label{f59}
\end{eqnarray}
where $E_1,E_2$ are constants. The state has no conical singularity at $A=0$, if $E_1=0$.
Because the conical singularity and the coordinate singularity are degenerated now due to Eq.(50),
we can say that in the quantisation of the black hole, the conical and the coordinate singularities are removed.

If we take a limit of $A\to \infty$, the Hamiltonian constraint becomes,
\begin{eqnarray}
\lim_{A\to \infty}H_S&=&-\frac{7}{8}A^2\frac{\partial^2}{\partial A^2}-\frac{Ac^2}{2} \nonumber \\
&\to &-\frac{7}{16}\frac{\partial^2}{\partial a'^2}-c^2=0 \label{f60}
\end{eqnarray}
Here, $a'=A^{1/2}=r$. 

This solution is,
\begin{eqnarray}
|\Psi (c,A\to \infty )\rangle =D_1\exp\bigg( i\frac{4}{\sqrt{7}}cA^{1/2}\bigg) +D_2\exp\bigg( -i\frac{4}{\sqrt{7}}cA^{1/2}\bigg) .\label{f61}
\end{eqnarray}
Here, $D_1,D_2$ are constants.
This function is different by $c$ is real or purely imaginary.
To satisfy the requirement that the averaged mass has the upper bound, we calculate the averaged mass in the infinity.
If $c$ is real,
\begin{eqnarray}
\langle M(c)\rangle \approx \int_0^{C}2\pi A^{1/2}\langle \hat{M}\rangle dA+\int_C^{\infty}2\pi (|D_1|^2+|D_2|^2)(A-c^2A^3)dA \nonumber \\
+\int_C^{\infty}2\pi (A-c^2A^3)(D_1D_2^*e^{i8cA^{1/2}/\sqrt{7}}+D_1^*D_2e^{-i8cA^{1/2}/\sqrt{7}})dA \label{f62}
\end{eqnarray}
Here, $C$ is some large number $(C\gg 1)$.
In this case, the second term of the r.h.s. diverges to $-\infty$ and the third term of the r.h.s. does not converges.
The first term of the right hand side is bounded from above, because we remove singularity.
So if $c$ is real the contradiction is appeared because of the behaviour of the averaged mass.
If $c$ is purely imaginary, the state becomes, as
\begin{eqnarray}
|\Psi (c,A\to \infty )\rangle =D_1\exp\bigg( -\frac{4}{\sqrt{7}}|c|A^{1/2}\bigg) +D_2\exp\bigg( \frac{4}{\sqrt{7}}|c|A^{1/2}\bigg)  . \label{f63}
\end{eqnarray}
The constant $D_2$ is zero, because the norm of the state should converge.
Then averaged value of the mass is,
\begin{eqnarray}
\langle M(c)\rangle \approx \int_0^{C}2\pi A^{1/2}\langle \hat{M}\rangle dA+\int_C^{\infty}2\pi |D_1|^2(A+|c|^2A^3)\exp\bigg( -\frac{8}{\sqrt{7}}|c|A^{1/2}\bigg) dA. \label{f64}
\end{eqnarray}
We know the second term of the r.h.s. converges because of exponential term and $\langle M(c)\rangle$ is bounded from above. 
If $c$ is purely imaginary, the averaged mass $\langle M(c)\rangle$ is always positive, so there does not appear any contradiction.
From above discussions, we can say that $c$ is purely imaginary.

Expanding the $|\Psi ^{4(5)}(c,A)\rangle $ state to the $|\Psi^{5(4)}(c,f,A)\rangle $ state, we simply assume $C_1$ is a function of $f$. 
Although $f$ and $A$ are related in ${\cal H}_{4(5)}$, however, we assume that $f$ and $A$ are independent in ${\cal H}_{5(4)}$.
Using the simplified 3+1 Hamiltonian constraint and the parameter relation in 
${\cal H}_{4(5)}$, we can simplify $m\hat{H}_S$ as
\begin{eqnarray}
m\hat{H}_S=-f^2\frac{\partial^2}{\partial f^2}+\frac{4}{7}\frac{c-cf}{f^{1/2}}=0. \label{f65}
\end{eqnarray}
This constraint equation has always solutions which we symbolically write as $C_1(c,f)$. 
Then the states of the expanded Hilbert space are symbolically written as $ |\Psi^{5(4)}(c,f,A)\rangle =C_1(c,f) F(c,A) $.
The existence of the expanded Hilbert space certificates the technique of up-to-down method. 
\section{Conclusion and discussions}\label{sec4}
The "up-to-down" method introduced in this paper looks as an interesting and powerful method to quantize important models in quantum gravity as a static solution.
Otherwise the Hilbert space may not contain all ${\cal H}_4$ state in this method, 
the subset of physical quantum state is exist and this state is easy to find.
We can easily find the static solutions of the Hamiltonian constraint by this method.
From now on there are no static restriction in the quantum gravity.
However, we find the static restriction in the quantum gravity.
Usually static solution is easier to solve than to solve the dynamical solutions.
We can say that it is same in the quantum gravity.

We studied in details  three mini-superspace models i.e. a Friedmann model, an anti-orthogonal model, and a black hole model.
In the Friedmann model we could check the up-to-down method is applicable.
In the anti-orthogonal model we could derive only a trivial solution. 
However, the combination of the parameters of the expanded state is non-trivial.
The major progress within the mini-superspace models seems to come from the black hole quantization.
The up-to-down method allows to transform a two parameter partial differential equation into an ordinary second order differential equation.

And we comment on the enlargement.
The three example of the mini-superspace are multiplication enlargement which are consistent of the measure of the projection.
So we can say we can find at least one enlargement whose projection is non-zero.

We succeeded in the quantization of the Schwarzschild black holes.
And we found the disappear of the conical and coordinate singularities at the quantum level. 
From the additional constraint, we could determine the explicit form of the mass operator and how to calculate the averaged mass. 

The up-to-down method introduced in this paper is technically introduced,
and the artificial Hilbert space is not the usual quantum gravity space.
Although there may be a relation to the higher dimensional quantum gravity Hilbert space,
we do not comment on it in this paper, because it is beyond the present work.

\begin{acknowledgments}
We would like to acknowledge A.Carlini and A.Hosoya for comments and discussions. 
\end{acknowledgments}

\end{document}